# The impact of outgassing of $CO_2$ and prior calcium precipitation to the isotope composition of calcite precipitated on stalagmites. Implications for reconstructing climate information from proxies.

Wolfgang Dreybrodt*,[1] and Jens Fohlmeister[2,3,4]

* Corresponding author: dreybrodt@t-online.de
[1]Faculty of Physics and Electrical Engineering, University of Bremen, Germany
[2]Potsdam Institute for Climate Impact Research, Telegrafenberg, 14473 Potsdam, Germany
[3]GFZ German Research Centre for [3]Geosciences, Section 'Climate Dynamics and Landscape Development', 14473 Potsdam, Germany
[4]German Federal Office for Radiation Protection (BfS), Koepenicker Allee 120, 10318, Berlin, Germany

**Abstract**

Degassing of $CO_2$ and precipitation of calcite to the surface of stalagmites can strongly impact isotope signals imprinted into the calcite of these speleothems. Here, we show that in all the variety of conditions occurring in nature only two distinct types of degassing exist. First, when a thin film of calcareous solution comes in contact to cave air lower $p_{CO2}$ value than that of the aqueous $CO_2$ in the water, molecular $CO_2$ escapes by physical diffusion in several seconds. In a next step lasting several ten seconds, pH and DIC in the solution achieve chemical equilibrium with respect to the $CO_2$ in the cave atmosphere. This solution becomes supersaturated with respect to calcite. During precipitation for each unit $CaCO_3$ deposited one molecule of $CO_2$ is generated and escapes from the solution. This precipitation driven degassing is active during precipitation only. We show that all variations of out gassing proposed in the literature are either diffusive outgassing or precipitation driven degassing and that diffusive outgassing has no influence on the isotope composition of the $HCO_3^-$ pool and consequently on that of calcite. Its isotope imprint is determined solely by precipitation driven degassing in contrast to most explanations in the literature. We present a theoretical model of $\delta^{13}C$ and $\delta^{18}O$ that explains the contributions of various parameters such as changes in temperature, changes of $p_{CO2}$ in the cave atmosphere, and changes in the drip intervals to the isotope composition of calcite precipitated to the apex of the stalagmite. We use this model to calculate quantitatively changes of $\delta^{13}C$ and $\delta^{18}O$ observed in field experiments (Carlson et al., 2020) in agreement to their experimental data. We also apply our model to prior calcite precipitation (PCP) in the field as reported by Mickler et al. (2019). We discuss how PCP influences isotope composition signals. In summary, we present a transparent method based on few commonly accepted equations that allows calculation of the isotope composition $\delta^{13}C$ and $\delta^{18}O$ of $CaCO_3$ under various temperatures, $p_{CO2}$ in the cave air, degrees of PCP, and concentrations $c_{Ca}^o$ of the water entering the cave.

## 1. Introduction

$\delta^{13}C$ and $\delta^{18}O$ records of calcite in stalagmites are used as important proxies for past climate. However, the climate signal and isotope variations, introduced by in cave processes, interfere with each other. Degassing of $CO_2$ from a $CaCO_3$-$CO_2$-$H_2O$ solution plays an important role to the isotope composition of $^{13}C$, $^{18}O$ in the calcite precipitated to stalagmites from this solution. For a climate interpretation of these proxies detailed understanding of the in cave processes is of utmost importance as highlighted in many studies (e.g., Frisia et a al., 2011, Yina, 2020; Yan et al., 2020).

Statements, which stress the importance to the rate of $CO_2$ degassing by the $p_{CO2}$ gradient between the drip water and the cave atmosphere, which is strictly speaking rather a $p_{CO2}$ difference than a gradient to the rate of $CO_2$ degassing, are often found in the literature (e.g., Lachniet, 2009; Wang et al., 2018). But they imply that isotope signals are imprinted to the calcite by outgassing of $CO_2$ that is driven by the $p_{CO2}$ gradient between drip water and cave air.

"*When calcite is deposited, $CO_2$ degassing is mainly controlled by the gradient of $CO_2$ concentration between drip water and cave air*" (Wang et al., 2018).

"*As degassing of $^{13}C$-depleted $CO_2$ progresses, the reservoir of $HCO_3$ in the water film will become isotopic heavier and deposited calcite thus will be enriched in heavy $^{13}C$.*" (Pua et al., 2016)

Similar versions for the assumed influence of $CO_2$-difference driven degassing of $CO_2$ on the $\delta^{13}C$ of dissolved inorganic carbon and precipitated $CaCO_3$ are given frequently in the literature (e.g., Mickler et al. 2006; 2019; Tremaine et al., 2011, Affek and Zaarur, 2014, Linge et al., 2009). Those phrases pose doubt on a detailed understanding of the $CO_2$ degassing and $CaCO_3$ precipitation processes. For instance, no exact definition of the gradient is given in these papers. The lack of understanding is replaced by many types of outgassing. Indistinct terms like "*forced, enhanced, rapid, intense, slow, fast, minimal, increased,* *equilibrium and progressive outgass*ing" are scattered throughout the literature without clear definitions of their meaning. One can summarize the idea underlying all these statements as follows: During precipitation of calcite, $CO_2$ degassing is governed solely by the difference, "gradient" between $p_{CO2}$ in the water and that in the cave atmosphere. This outgassing process drives calcite precipitation. Due to preferential degassing of the light $CO_2$ isotopologue DIC in the solution is enriched in the heavy isotope.

**This concept is not completely correct. In some parts it is misleading.**

However, there are few works that give alternative views (Johnston et al., 2013):

"*As the drip water enters the cave, $CO_2$ picked up from the soil degasses by diffusion* caused *by a typically lower $p_{CO2}$ in the cave atmosphere than that of the emergent drip water. On completion of degassing to equilibrium with the cave environment, the pH has increased (to approximately 8.5) and the majority (> 95 %) of the carbonate in solution is in the bicarbonate ($HCO_3^-$) form. This determines the stable isotope values of the DIC but has no influence on speleothem $\delta^{13}C$ and $\delta^{18}O$ values since calcite precipitation starts after the establishment of supersaturation*."

Recent work has confirmed such findings (Dreybrodt, 2019, Guo and Zhou, 2019, Yan et al., 2020, Drăgușin et al., 2020). These messages, however, so far have not been commonly perceived in the literature.

Nevertheless, there is general agreement that understanding of processes that happen inside caves are of high significance in the interpretation of stalagmite climate proxies. However, not only knowledge about processes on the stalagmite surface is required, but also about processes, that occur before a drop impinges on the top of a speleothem. One important process is termed prior calcite precipitation (PCP). Usually this term is used when cave water precipitates $CaCO_3$ before it drips to the speleothem top. This process is regarded in many studies as a major driver for changes in the isotopic composition of speleothems (e.g., Sherwin and Baldini, 2011; Riechelmann et al., 2013; Fohlmeister et al., 2017; 2020). However, the effect of PCP on growth rate and the stable $^{13}C$ and $^{18}O$ isotope composition was not yet determined quantitatively.

In this work, we provide a comprehensive description of outgassing and PCP and their influence to $CaCO_3$ precipitated on the speleothem surface. We summarize the physics and chemistry of the processes that cause degassing, precipitation of calcite and evolution of isotope signals from a thin water film supersaturated with respect to calcite, as they exist on speleothem surfaces. While most of this information is drawn from earlier work (Hansen et al., 2013, 2019, Dreybrodt and Scholz, 2011, Dreybrodt, 2019a, b, c) we provide the first quantitative description of the influence of PCP on the isotopic composition of speleothem $CaCO_3$.

From this information, we construct a model to calculate changes of $\delta^{13}C$ and $\delta^{18}O$ observed in field experiments (Carlson et al., 2020) in agreement to their experimental data. We apply our model also to prior calcite precipitation (PCP) in the field as reported by Mickler et al. (2019). Furthermore, we discuss the isotope composition of calcite precipitated under various climatic conditions.

**2. Processes determining the isotope composition of calcite precipitated on the surface of stalagmites**

**2.1 Diffusive outgassing of aqueous CO₂ and evolution until start of precipitation of calcite**

There are three main pools of carbon that interact with each other: 1) Aqueous $CO_2$ molecules in the solution, 2) $CO_2$ molecules in the cave atmosphere, and 3) $HCO_3^-$ ions in the solution. A further pool of $CO_3^{2-}$ can be neglected at pH-values as they exist under typical cave conditions.

The drop impinging to the stalagmite forms a thin layer with depth, a. In the first step aqueous $CO_2$ molecules escape by molecular diffusion from the solution to the atmosphere. Since the abundant and rare isotopes do not interact (Dreybrodt and Scholz, 2011) at the end of outgassing each species is in chemical equilibrium with the corresponding species in the cave atmosphere. In chemical and isotope equilibrium the concentration, $c_{eq}^{CO2,i}$, of each isotopologue, i, of aqueous $CO_2$ in the water is related to the corresponding partial pressure, $p_{cave}^{CO2,i}$, of each isotopologue, i, of $CO_2$ in the cave atmosphere by Henry's law, $c_{eq}^{CO2,i} = K_H^i \cdot p_{cave}^{CO2,i}$. $K_H^i$ is Henry's constant. Note that this way, aqueous $CO_2$ is in isotope equilibrium with the $CO_2$ in the atmosphere. But it is not in isotope equilibrium with $HCO_3^-$. This disequilibrium causes isotope exchange of atmospheric $CO_2$ with $HCO_3^-$ that proceeds via aqueous $CO_2$ until $HCO_3^-$, aqueous and atmospheric $CO_2$ are in isotope equilibrium with each other. This process however is slow. Under conditions as they exist in caves the exponential exchange time is on the order of ten thousand seconds (Dreybrodt et al., 2016, Dreybrodt, 2017). Therefore its contribution to the isotope composition of calcite can be neglected (Dreybrodt and Romanov, 2016).

For water layers with a depth, a, of several tenths of a millimeter outgassing is fast and takes a few seconds in agreement to the theoretically predicted time, $\tau_{diff}$.

$$\tau_{diff} = 4a^2 / (\pi^2 D) \qquad (1)$$

where D = 1.26·10⁻⁵ cm²s⁻¹ at 10°C, 1.67·10⁻⁵ cm²s⁻¹ at 20°C, and 2.17·10⁻⁵ cm²s⁻¹ at 30°C, is the diffusion constant of aqueous $CO_2$ (Jähne et al., 1987). Note that the time constant, $\tau_{diff}$, for outgassing is independent of the difference between the $p_{CO2}$ in the solution and in the cave atmosphere. During the first step of diffusive outgassing the solution remains undersaturated with respect to $CaCO_3$ and calcite cannot precipitate. Therefore, the $Ca^{2+}$- concentration remains constant. Throughout this work, we refer to precipitation of calcite. All the results, however, are qualitatively valid for aragonite precipitation as well. The $HCO_3^-$ - concentration is tied to the $Ca^{2+}$- concentration by electro neutrality $[HCO_3^-] + [OH^-] = 2[Ca^{2+}] + [H^+]$. For changes of pH from 7 to 8 as they are common, $[H^+]$ and $[OH^-]$ change by about $10^{-7}$ mol/L, approximately four orders of magnitude smaller than $[Ca^{2+}]$. Therefore $[HCO_3^-] = 2[Ca^{2+}]$ is valid.

Consequently, within this approximation, [$HCO_3^-$] must stay constant as well. Thus, during diffusive outgassing of $CO_2$ the isotope composition of the $HCO_3^-$ pool stays unaffected.

Precipitation starts when the saturation index, $SI_{calcite}$, reaches a value of about 0.7. Reaching such a SI value takes the time of equilibration of the DIC species, $\tau_{eq}$, which is on the order of several ten seconds (Zeebe et al., 1999). Hansen et al. (2013) have verified this experimentally at 25°C. They measured the pH value of a water film of a solution flowing down an inclined limestone plate immediately after outgassing analogously to the situation of a stalagmite. They found an equilibration time, $\tau_{eq}$ = 12 ±4 s, in satisfactory agreement to the theoretical prediction by Zeebe et al. (1999). One should note that $\tau_{eq}$ decreases with increasing temperature and increasing pH (Dreybrodt and Romanov, 2016).

During the time, $\tau_{eq}$, of isotopic and chemical equilibration between dissolved $CO_2$ and $HCO_3^-$ (via $H_2CO_3$) until start of precipitation of calcite the amount of $H_2CO_3$ converted to $CO_2$ by the instantaneous reaction $HCO_3^- + H^+ \rightarrow H_2CO_3$ is equal to the change of the $H^+$ concentration of about $10^{-7}$ mol/L. This is small in comparison to the concentration of aqueous $CO_2$ of about $2 \cdot 10^{-4}$ mol/L at a $p_{CO_2}$ of 0.0004 atm. We explain this in detail in the Appendix. Therefore, during this stage the system can be regarded as closed with respect to aqueous $CO_2$ and $HCO_3^-$. Consequently, the isotope compositions, $\delta^{13,18}DIC$, of dissolved inorganic carbonates, DIC, must stay constant as well. Furthermore, during equilibration the concentrations of $CO_2$ and $HCO_3^-$ also remain unaltered. $c_{CO_2}$ is fixed to $p_{CO_2}$ in the cave atmosphere and the concentration of $HCO_3^-$ is tied to the calcium concentration, $c_{Ca}$, by electro neutrality. Therefore, we get $c_{CO_2}(out) = c_{CO_2}(eq)$ and $c_{HCO3}(out) = c_{HCO3}(eq)$ where out refers to the state after outgassing.

The isotope compositions for both oxygen and carbon of DIC after outgassing (out) and after equilibration (eq) are given by

$$\delta^{13,18}DIC(out) = \frac{\delta^{13,18}CO_2(out) \cdot c_{CO_2}(out) + \delta^{13,18}HCO_3(out) \cdot c_{HCO3}(out)}{c_{CO_2}(out) + c_{HCO3}(out)} \quad (2)$$

$$\delta^{13,18}DIC(eq) = \frac{\delta^{13,18}CO_2(eq) \cdot c_{CO_2}(eq) + \delta^{13,18}HCO_3(eq) \cdot c_{HCO3}(eq)}{c_{CO_2}(eq) + c_{HCO3}(eq)} \quad (3)$$

As already mentioned, $\delta^{13,18}HCO_3(out)$ is equal to the initial isotope composition $\delta^{13,18}HCO_3(initial)$ in the water prior to degassing. Because the system is closed with respect to carbonate and $CO_2$ during isotopic equilibration, the oxygen and carbon isotope compositions, $^{13,18}\delta_{DIC}$, of dissolved inorganic carbonate, DIC, must stay constant.

$$\delta^{13,18}DIC(out) = \delta^{13,18}DIC(eq) \quad (4)$$

Furthermore, the concentrations, $c_{CO_2}$, of $CO_2$ and, $c_{HCO3}$, of $HCO_3^-$ must also remain unaltered because $c_{CO_2}$ is fixed by Henrys law with $p_{CO_2}$ of the cave air $CO_2$ and $c_{HCO3}$ by the condition of electro neutrality. Combining eqns. 2 to 4 we obtain

$$\delta^{13,18}HCO_3(out) - \delta^{13,18}HCO_3(eq) = \left[\delta^{13,18}CO_2(out) - \delta^{13,18}CO_2(eq)\right] \cdot c_{CO_2}(out)/c_{HCO3}(out) \quad (5)$$

During equilibration with the $HCO_3^-$ pool that needs several ten seconds in comparison to the time of outgassing of only a few seconds, the $CO_2$-pool in the solution is in chemical and isotope equilibrium with the $CO_2$ in the atmosphere because each $CO_2$ isotopologue in the solution by diffusion approaches equilibrium with its counterpart in the atmosphere independent of each other. Therefore, $\delta^{13,18}CO_2(out) = \delta^{13,18}CO_2(eq)$ and consequently

$$\delta^{13,18}HCO_3(eq) = \delta^{13,18}HCO_3(out) = \delta^{13,18}HCO_3(initial) \qquad (6)$$

for the isotope composition of the $HCO_3^-$ pool. This means, that there is no impact of diffusion driven degassing and equilibration on the $HCO_3^-$ pool. Consequently no impact occurs to the calcite precipitated later on in contrast to current statements in the literature.

Hansen et al. (2019) have given experimental proof. They measured pH, electrical conductivity, $\delta^{13}C$, and $\delta^{18}O$ of DIC for a solution flowing on a limestone plate analogously to a stalagmite. The water with pH = 6.7 and Ca-concentration of 5 mmol/L drops to the plate. When it reaches the end of the plate outgassing to a $p_{CO2}$ = 0.001 atm is completed and the pH has increased to 8.0 documenting outgassing of $CO_2$ and chemical equilibration. From $\delta^{13}C$, $\delta^{18}O$, and pH at the input and at the output after degassing and equilibration they have calculated the $\delta^{18}O$ value of the $HCO_3^-$ pool. They found that both δ-values, at the entrance and at the exit of the plate were equal within the accuracy of their experiment. For $^{18}O$ the authors report a δ-value of -7.6‰ before outgassing and -7.2‰ after outgassing. For $^{13}C$ the corresponding values are less accurate and we estimated from their data -34.3‰ and -35∓1‰.

They state: "*degassing of $CO_2$ and establishment of supersaturation with respect to calcite on the upper plate does not affect the $\delta^{18}O$ value of the $HCO_3^-$.*".

To understand the temporal evolution of the concentration of DIC and its isotope δ-values one has to consider the temporal evolution of all species in DIC. The concentration, $c_{DIC}$, of DIC is the sum of the concentration of aqueous $CO_2$, $c_{CO2}$ and the concentration of $HCO_3^-$, $c_{HCO3}$. At pH about 8 the concentration of $CO_3^{2-}$ can be neglected and $HCO_3^-$ is more than 95% of DIC.

In the first step of degassing the temporal evolution of $c_{CO2}$ is given by

$$c_{CO2}(t) = (c_{CO2}^0 - c_{CO2}^{eq})\exp(-t/\tau_{diff}) + c_{CO2}^{eq} \qquad (7)$$

$c_{CO2}^0$ is the concentration in the water at the drip site when it falls to the stalagmite at time zero. $c_{CO2}^{eq}$ is the concentration in equilibrium with the $CO_2$ in the cave atmosphere, and $\tau_{diff}$ is the exponential time constant for degassing (Hansen et al., 2013, Dreybrodt, 1988).

For later times, $\tau_{diff}$ < t < $\tau_{diff}$ +3·$\tau_{eq}$, the $HCO_3^-$ pool achieves chemical and isotope equilibrium with the pool of remaining aqueous $CO_2$ (Hansen et al., 2013, Zeebe et. al., 1999). The only reaction that can affect the isotope composition of the $HCO_3^-$ pool is isotope exchange with the $CO_2$ in the atmosphere. This reaction, as already stated needs times of ten thousands seconds. Therefore, during the time of equilibration ($\tau_{eq} \cong 40\ s$) and subsequent precipitation (Dreybrodt et al., 2017, Dreybrodt and Romanov, 2016, Dreybrodt, $\tau_{prec} \cong 400\ s$) isotope changes in the $HCO_3^-$ pool by exchange with the $CO_2$ in the atmosphere are small and can be safely be neglected. For the oxygen isotopologues exchange with oxygen in the water is active on time scales of several ten thousands seconds (Beck et al., 2004). This reaction can also be neglected at the time scale of precipitation.

In the following, we discuss the temporal evolution of $^{13}C$ isotopologues. These results are also valid for the oxygen carbonate isotopologues because exchange with the water is slow. $\delta^{13}CO_2(t)$, in the first step of outgassing is determined by molecular diffusion. Generally the δ-value (here given as small numbers and not in $^O/_{OO}$) is related (Dreybrodt et al., 2016) to the concentration of the rare isotope by

$$\delta(t) - \delta(0) = \left( c^{rare}(t) / c^{rare}(0) - 1 \right) \tag{8}$$

we get

$$\delta^{13}CO_2(t) = (\delta^{13}CO_2^0 - \delta^{13}CO_2^{eq}) \exp(-t / \tau_{diff}) + \delta^{13}CO_2^{eq} \tag{9}$$

$\delta^{13}CO_2^{eq}$ Is the isotope composition in chemical and isotope equilibrium with the cave atmosphere. After the time t = 3τ$_{diff}$ out gassing can practically be regarded as completed. For later times t > 3τ$_{diff}$

$$\delta^{13}CO_2(t) = \delta^{13}CO_2^{eq} \tag{10}$$

At a $SI_{CaCO3}$ of about 0.7 precipitation starts and the calcium concentration, $c_{Ca}$, declines exponentially (Hansen et al., 2013) with time constant, τ$_{prec}$.

$$c_{Ca}(t) = (c_{Ca}^0 - c_{Ca}^{eq}) \exp(-t / \tau_{prec}) + c_{Ca}^{eq} \tag{11}$$

Because $c_{HCO3}$ is tied to $c_{Ca}$ by electro neutrality, $2 \cdot c_{Ca}(t) = c_{HCO3}(t)$ we have also (Hansen et al., 2013)

$$c_{HCO3}(t) = (c_{HCO3}^0 - c_{HCO3}^{eq}) \exp(-t / \tau_{prec}) + c_{HCO3}^{eq} \tag{12}$$

To keep an overview on the times, τ$_{prec}$, , τ$_{eq}$, and τ$_{diff}$ these times are listed for various temperatures in Table 1.

| Temperature °C | τ$_{diff}$ (s) | τ$_{eq}$ (s) * | τ$_{prec}$ (s) |
|---|---|---|---|
| 10 | 3.2 | 111 | 800 |
| 15 | 2.8 | 66 | 570 |
| 20 | 2.4 | 40 | 360 |
| 30 | 1.8 | 16 | 190 |

*Table 1: Temperature dependence of τ$_{diff}$, τ$_{eq}$, and τ$_{prec}$ for a layer depth, a, of 0.01cm. τ$_{eq}$ is almost independent of film depth a. * Johnson (1982), Dreybrodt (2017)*

At that moment, a comment is necessary. We deal with three processes that we have considered to happen consecutively. This assumption is true only if the time, $\tau_{diff}$, needed for degassing is small in comparison to the equilibration time, $\tau_{eq}$. During degassing, as soon as the $CO_2$-concentration drops equilibration starts. If $\tau_{eq} \gg \tau_{diff}$ the progress of equilibration during the short time $\tau_{diff}$ is small because its completion needs the much longer time $\tau_{eq}$. Furthermore, during equilibration as soon as the saturation index, $SI_{CaCO3} > 0$, calcite can precipitate. However, because $\tau_{eq} \ll \tau_{prec}$ one can neglect the contribution of this progress of precipitation during the short time $\tau_{eq}$. In summary, the three processes can be considered as consecutive only if $\tau_{diff} \ll \tau_{eq} \ll \tau_{prec}$. This is the case in the system $CaCO_3$-$H_2O$-$CO_2$ as the time constants differ each by at least one order of magnitude. (Dreybrodt and Scholz, 2011, Dreybrodt et al., 2016, Dreybrodt, 2017). We have calculated the temporal evolution of the concentrations $c_{CO2}(t)$, $c_{HCO3}(t)$, and the isotope composition of $CO_2$, $HCO_3^-$, and DIC. The results are shown in Figure 1. The water dripping to the stalagmite is in chemical equilibrium with respect to a $p_{CO2} = 0.02$ atm at a temperature of 20°C with an initial concentration $c_{CO2}^0 = 0.78\ mmol/L$, $c_{HCO3}^0 = 4.47\ mmol/L$, and $c_{Ca}^0 = 2.24\ mmol/L$. Both species are in chemical and isotope equilibrium with each other. The δ-values with respect to VPDB are $\delta^{13}CO_2^0 = -21.5‰$ and $\delta^{13}HCO_3^0 = -12‰$. The cave atmosphere contains 0.0004 atm $CO_2$ with $\delta^{13}CO_2^{cave} = -8‰$. After completion of outgassing the aqueous $CO_2$ has a δ-value $\delta^{13}CO_2^{eq} = -9‰$ in equilibrium with the $CO_2$ in the cave atmosphere.

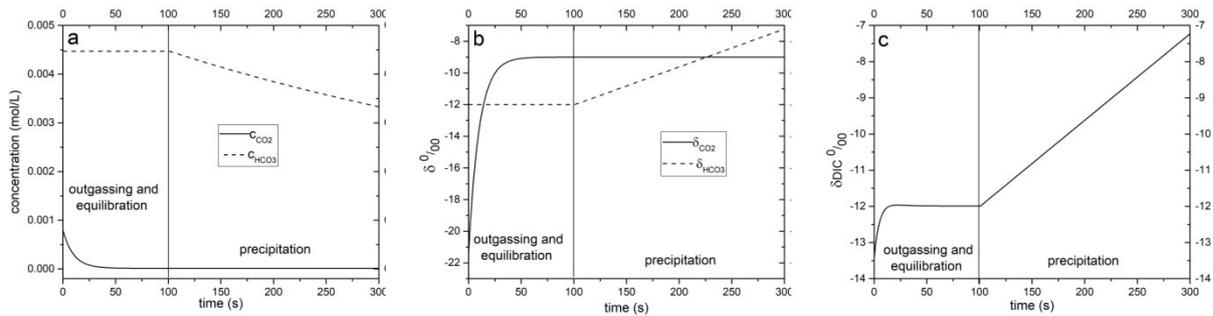

*Fig.1: Temporal evolution of the concentrations and δ-values: a) temporal evolution of the concentrations $c_{CO2}(t)$ and $c_{HCO3}(t)$. b) Temporal evolution of $\delta^{13}CO_2(t)$ and $\delta H^{13}CO_3(t)$. c) Temporal evolution of isotope composition $\delta^{13}DIC(t)$. All δ are in VPDB. Changes in $c_{CO2}(t)$ and $\delta^{13}CO_2(t)$ are restricted to the time of outgassing and equilibration, whereas changes in $c_{HCO3}(t)$ and $\delta^{13}HCO_3(t)$ take place thereafter during precipitation of calcite.*

Due to outgassing, within 20 s the concentration $c_{CO_2}(t)$ drops from 0.78 mmol/L to 15.6 μmol/L (Fig.1a) and then remains constant. $\delta^{13}CO_2$ rises from its initial value of -21.5‰ to -9‰ in equilibrium with the $CO_2$ in the cave atmosphere and then remains constant (Fig1b). The concentration $c_{HCO3}(t)$ of bicarbonate remains constant until precipitation starts after 100 s and then declines exponentially with a time constant $\tau_{prec} = 500\ s$ (Fig.1a). During precipitation, aqueous $CO_2$ is in equilibrium with the $p_{CO2}$ of the cave atmosphere.

It is important to stress the conclusion: There is no gradient of $p_{CO2}$ during precipitation of calcite as often stated in the literature. There is a difference of $p_{CO2}$, which, however exists only during the first step of diffusion controlled outgassing that reduces this difference to zero.

## 2.2. Outgassing driven by precipitation of calcite.

Precipitation rates of $CaCO_3$ from supersaturated solutions in the $H_2O$ - $CO_2$ - $CaCO_3$ system are controlled by three processes acting simultaneously: 1) the kinetics of precipitation at the mineral surface expressed by the PWP-equation (Plummer et al., 1978), 2) diffusion mass transport of the reaction species involved to and from the mineral surface, and 3) the slow kinetics of the overall reaction $HCO_3^- + H^+ \rightarrow CO_2 + H_2O$. To obtain precipitation rates under these various processes, which control the rates; one has to solve the transport-reaction equations which take into consideration these three mechanisms. This has been done by Buhmann and Dreybrodt (1985) and was later verified by Kaufmann and Dreybrodt (2007). Their results show that the precipitation rates $F$ are given by a linear relation

$$F = \alpha \cdot (c_{Ca} - c_{Ca}^{eq}) \tag{13}$$

F is in mmol/(cm$^2$ s). $c_{Ca}^{eq}$ is the Ca equilibrium concentration with respect to calcite and pco2 in the cave atmosphere. The constant, $\alpha$ (cm/s) depends on temperature by the relation $\alpha = (0.52+0.04T+0.004T^2)\cdot 10^{-5}$ cm/s and increases by about a factor of ten from 0°C to 25°C (Baker et al., 1998). T is temperature in °C. Furthermore it depends on the depth $a$ of the water layer (Baker et al., 1998). For water layer depths as they occur on stalagmites, however, a constant value can be applied. The theoretical results have been verified experimentally in the field (Baker et al., 1998) and in the lab (Hansen et al., 2013, Dreybrodt et al., 1997, Dreybrodt, 2012)

From eqn. 13, one obtains the temporal evolution of the Ca-concentration (Dreybrodt, 1988) as already stated in eqn. 11 and in eqn. 12

$$c_{Ca}(t) = (c_{Ca}^0 - c_{Ca}^{eq})\exp(-t/\tau_{prec}) + c_{Ca}^{eq} \tag{14}$$

Precipitation times, $\tau_{prec} = a/\alpha$, range between 1000 s and 100s for $\delta$ = 0.01 cm and T = 0°C and 25°C, respectively. $c_{Ca}(t)$ is the actual calcium concentration in mmol/cm$^3$ in the water film and $c_{Ca}^o$ is the initial concentration of calcium.

Due to the stoichiometry of the reaction $Ca^{2+} + 2HCO_3^- \rightarrow CaCO_3 + CO_2 \uparrow + H_2O$ for each unit of $CaCO_3$ deposited one molecule of $CO_2$ and two $HCO_3^-$ ions are removed from the solution by outgassing of one molecule of $CO_2$ and deposition of one unit of $CaCO_3$. This has also been shown experimentally by Dreybrodt (2019). Consequently we have

$$c_{HCO3}(t) = (c_{HCO3}^0 - c_{HCO3}^{eq})\exp(-t/\tau_{prec}) + c_{HCO3}^{eq} \tag{13}$$

The evolution of the isotope composition of $HCO_3^-$ remaining in the solution can be described by Rayleigh distillation (Dreybrodt and Scholz, 2011, Dreybrodt, 2016, Hansen et al., 2019).

After onset of precipitation for t< 0.3·$\tau_{prec}$ one gets by expansion to first order in t(Dreybrodt and Scholz, 2011, Dreybrodt, 2016)

$$\delta^{13}HCO_3(t) = \delta^{13}HCO_3^o - \varepsilon_{kin} \cdot t/\tau_{prec} \tag{14}$$

$\varepsilon_{kin}$ is a kinetic constant of isotope enrichment due to precipitation and has been defined theoretically (Dreybrodt, 2016, Dreybrodt and Scholz, 2011, Dreybrodt and Romanov, 2016, Dreybrodt, 2019) and observed experimentally (Hansen et al., 2019). After t = $\tau_{prec}$, 65 % of the calcite has been deposited and the linear behaviour of $\delta^{13}HCO3(t)$ bends over to a constant value. Fig. 1c depicts the temporal evolution of $\delta^{13}C$ in DIC and the $HCO_3^-$ pool. $\delta^{13}DIC(t)$ starts at a value of -13.5 ‰ and rises due to outgassing to -12 ‰. This is the isotope composition of the $HCO_3^-$ reservoir because now $HCO_3^-$ is abundant by more than 95%. Then, after onset of precipitation $\delta^{13}DIC$ (t) rises in the same way as $\delta^{13}HCO_3(t)$ (Fig.1c). In all panels, the time is limited to the range where the approximation of eqn. 10 is valid.

## 2.3. Influence of $p_{CO2}^{cave}$ to the isotope composition of the calcite precipitated

A change of $p_{CO2}$ in the cave atmosphere causes a change of, $c_{Ca}^{eq}$, the equilibrium concentration of calcium with respect to calcite and $CO_2$ by the relation $c_{Ca}^{eq} = K(T)\sqrt[3]{p_{CO2}}$ where K(T) is a constant depending on temperature, T via the mass action constants of the carbonate chemistry (Dreybrodt, 1988). By use of PHREEQC (Parkhurst and Appelo, 1999) we find K = -0.16·T+11.13. T is in °C and $c_{Ca}^{eq}$ in mmol/L (K = 9.5 at 10°C, 8.02 at 20°C, and 6.28 at 30°C) as a satisfactory approximation.

According to eq. 13, the precipitation rate, F, can be expressed as (Baker et al., 1998, Buhmann and Dreybrodt, 1985)

$$F = \alpha \cdot (c_{Ca} - K(T)\sqrt[3]{p_{CO2}}) \tag{15}$$

The precipitation rates decrease with increasing $p_{CO2}$. They are depicted in Fig. 2. The corresponding rates of $CO_2$ outgassing controlled by calcite precipitation are equal to the withdrawal rates of calcium from the solution by precipitation of calcite. During precipitation of calcite outgassing is determined by the precipitation rates and not by $p_{CO2}$ difference (gradient) between solution and cave atmosphere as stated many times in the literature.

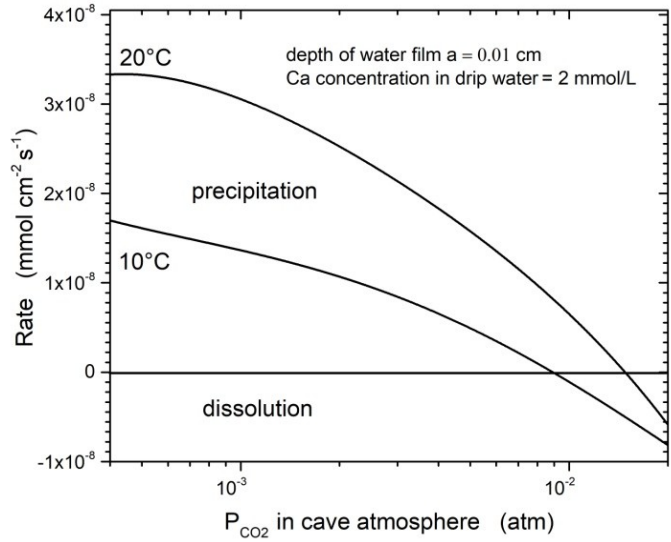

*Fig. 2. Initial precipitation rates $F = \alpha \cdot (c_{Ca}^0 - K(T)\sqrt[3]{p_{CO2}})$ in dependence of $p_{CO2}$ in the cave atmosphere at 10°C and 20°C. The horizontal line divides in the upper region of precipitation and the lower one of dissolution. At the intersections of this line and the curves, one can read the $p_{CO2}^{eq}$ in equilibrium with respect to calcite whereas $p_{CO2} > p_{CO2}^{eq}$ indicate dissolution at the stalagmite. All values $p_{CO2} < p_{CO2}^{eq}$ refer to precipitation.*

At that point it is important to realize that only outgassing caused by precipitation of calcite has an impact on the isotope composition of $HCO_3^-$ in the solution and consequently on the calcite precipitated. The influence of $p_{CO2}$ in the cave atmosphere to the isotope composition of the calcite deposited to the stalagmite results solely from the relation $c_{Ca}^{eq} = K(T)\sqrt[3]{p_{CO2}}$ in eqns. 18a, b, of Rayleigh distillation.

### 2.4 Isotope composition of the calcite precipitated

The change in isotope composition of calcite results from a Rayleigh fractionation process of $HCO_3^-$ in the stagnant water film that is replaced by each new drop. Within the time, $T_{drip}$, between two drops $\delta^{13}C$ and $\delta^{18}O$ in the $HCO_3$ pool in the solution increase and consequently in the calcite deposited. In the case of classical Rayleigh distillation only one fractionation constant, ε, is sufficient to describe the evolution of the isotope compositions by

$$R(c(t))/R(c(0)) = (c(t)/c(0))^\varepsilon = (1+\delta(c(t)))/(1+\delta(c(0))) \tag{18a}$$

R is the isotope ratio and δ is the isotope composition of $HCO_3^-$ in the water. δ and ε are in small numbers, not in ‰. In the following, we are interested only in the change of R after the water has stayed at the stalagmite during the drip time, $T_{drip}$, until a new drop replaces it. Therefore we take R(c(0)) = 1 and δ(c(0)) = 0. c(0) is the initial concentration of $HCO_3^-$ when precipitation starts. Its concentration c(t) is given by eqn. 14.

However, applying the classical Rayleigh approach to calcites in nature has shortcomings. The theoretical basis for Rayleigh distillation is chemical and isotope equilibrium of all reacting agents. Most of Earth-surface calcites, however, precipitate out of isotopic equilibrium (Daeron et al., 2019). To take account of this one regards the fractionation, ε, as kinetic, $\varepsilon^{kin}$, and one uses eqn. 18a with this factor. $\varepsilon^{kin}$ has been determined in cave analogous experiments by Hansen et al. (2019) for a variety of temperatures, $p_{CO2}$ in the atmosphere, and Ca-concentrations of the water dripping to the speleothem.

Alternatively, a kinetic approach has been suggested (Dreybrodt, 2008, Dreybrodt and Scholz, 2011, Dreybrodt, 2016) termed the extended Rayleigh equation. Here, the isotopologues react independently of each other. In the reaction, $F = \alpha(c_{Ca} - c_{Ca}^{eq})$ the reaction constant, α, differs for the heavy and the light isotopologues. The isotope fractionation $\alpha^{kin}$ is defined by the ratio $\alpha_{heavy}/\alpha_{light} < 1$, whereby one assumes that the light isotope reacts faster than the heavy one. Furthermore, one accounts for different equilibrium concentrations with respect to calcite of the light and heavy isotopes.

$c_{Ca,heavy}^{eq} / c_{Ca,light}^{eq} = \gamma \cdot c_{Ca,heavy}(0) / c_{Ca,light}^{eq}(0)$. Here, $c_{Ca,heavy}(0) / c_{Ca,light}^{eq}(0)$ is the isotope ratio of the initial solution dripping to the stalagmite and the fractionation $\gamma \cong 1$. From this, one obtains the extended Rayleigh equation.

$$R(c_{Ca}(t))/R(c_{Ca}^0) = \left[\left(\frac{c_{Ca}(t) - c_{Ca}^{eq}}{c_{Ca}^0 - c_{Ca}^{eq}}\right)^{\alpha_{kin}} \cdot \left(1 - \gamma \frac{c_{Ca}^{eq}}{c_{Ca}^0}\right) + \gamma \frac{c_{Ca}^{eq}}{c_{Ca}^0}\right] \cdot \frac{c_{Ca}^0}{c_{Ca}(t)} \tag{18b}$$

Using the data of Hansen et al. (2019) Dreybrodt (2019) has determined the values of $\alpha^{kin}$ and γ. This gives evidence that the extended Rayleigh approach is more appropriate. Nevertheless, as shown in the Appendix the results of both models are very similar. We therefore, for convenience use the classic Rayleigh distillation in this work. We stress, however, that the kinetic approach seems to be more appropriate. Here, we employ the classical approach because the community is used to it and the results within the limits of uncertainty are undistinguishable.

The average value of the isotope ratio, R, in the calcite deposited during one drip interval is given by (Dreybrodt and Scholz, 2011).

$$\bar{R} = \frac{\int_{c_{Ca}^0}^{c_{Ca}(Tdrip)} R(c_{Ca}) \cdot dc_{Ca}}{\int_{c_{Ca}^0}^{c_{Ca}(Tdrip)} dc_{Ca}} + \varepsilon_{CaCO3/HCO3} = \frac{R(c_{Ca}^0) \int_{c_{Ca}^0}^{c_{Ca}(Tdrip)} (c_{Ca}/c_{Ca}^0)^\varepsilon \cdot dc_{Ca}}{\int_{c_{Ca}^0}^{c_{Ca}(Tdrip)} dc_{Ca}} + \varepsilon_{CaCO3/HCO3} =$$

$$= \frac{R(c_{Ca}^0) \int_0^{Tdrip} \alpha \cdot (c_{Ca}(t)/c_{Ca}^0)^\varepsilon \cdot (c_{Ca}(t) - c_{Ca}^{eq}) \cdot dt}{\int_0^{Tdrip} \alpha \cdot (c_{Ca}(t) - c_{Ca}^{eq}) \cdot dt} + \varepsilon_{CaCO3/HCO3} \tag{19}$$

$c_{Ca}(t)$ is the actual concentration, given by eqn. 14. $c_{Ca}^0$ is the initial calcium concentration of the drop when it falls to the stalagmite at time zero. $c_{Ca}(T_{drip})$ is the concentration in the water film after the drip interval $T_{drip}$. Its minimal value is $c_{Ca}^{eq}$. $\varepsilon_{CaCO3/HCO3}$ is the fractionation constant between $HCO_3^-$ and calcite. Since in this work we deal with differences $\delta(t) - \delta(0)$ the results are independent of $\varepsilon_{CaCO3/HCO3}$ Therefore, for convenience we use $\varepsilon_{CaCO3/HCO3} = 0$ throughout the paper.

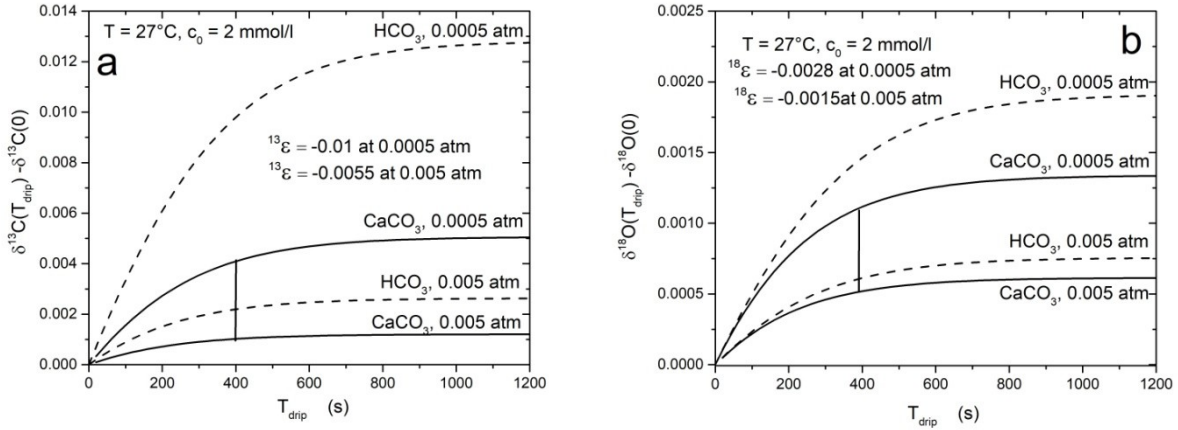

Figure 3: $\delta^{13}C$ and $\delta^{18}O$ in the $HCO_3^-$ pool and in the calcite deposited from a water layer of depth, $\delta$ = 0.01 cm after drip time $T_{drip}$. The vertical lines give the difference between high and low $p_{CO2}$. $c_{Ca}^0$ = 2mmol/L, kinetic fractionation constants $^{13}\varepsilon$ = -0.010 and $^{18}\varepsilon$ = -0.003

Using eqn. 18a and numerical integration (Origin, www.OriginLab.com) we obtain the result for $\delta^{13}C$ and $\delta^{18}O$ as shown in Fig. 3. $c_{Ca}^0$ = 2mmol/L, the kinetic fractionation constants $^{13}\varepsilon$ = -0.010 and $^{18}\varepsilon$ = -0.003 are taken from the experimental work of Hansen et al. (2019). Temperature in the cave is 27°C. It should be noted here that the precipitation rates of calcite for high $p_{CO2}$ are only one third of those at low $pCO_2$ (see Fig. 2).

### 3. Impact of drip time and $p_{CO2}$ in the cave atmosphere to the isotope composition of calcite.

Seasonal changes in the cave ventilation regime are responsible for variations of $p_{CO2}$ in the cave atmosphere. In downward leading caves the summer season $p_{CO2}$ is usually higher than during the cold season and hence precipitation rates are low (see Fig. 2). As a result, the increase in $\delta^{13}C$ and $\delta^{18}O$ of $HCO_3^-$ and precipitated calcite in the summer season is low compared to the winter time.

The vertical lines between the corresponding curves in Fig. 3 depict the change $\delta$ between high and low $p_{CO2}$. Carlson et al. (2020) have observed this in a cave. They report that at sites with low drip rates (drip site "Stumpy") and extended periods with high (0.005 atm) and low (0.0005 atm) cave-air $CO_2$ but constant temperature of 27°C throughout the year, isotope $\delta^{18}O$ exhibits shifts of about 1‰ in calcite precipitated to glass plates located below the drip site. For $\delta^{13}C$ they find values of about 2 ‰. We try to model this with our approach (Sec. 2) using values as observed for this drip site. At the drip site Stumpy drip interval, $T_{drip}$, is 400 s $\mp$ 200 s. For T = 27°C the kinetic constant is $\alpha = 5 \cdot 10^{-5}$ cms$^{-1}$. The depth of the water film is assumed to be a = 0.01 cm. Therefore, **$\tau_{prec}$** = 200 s ± 100 s.

Fig. 3 depicts also quantitatively the dependence of $\delta^{13,18}$ on the drip time for this case. The difference of 0.5 ‰ in $\delta^{18}O$ of the precipitated carbonate between high and low $p_{CO2}$ is indicated by the vertical lines at $T_{drip}$ = 400 s. This value is comparable with observation of values of about 1 $\mp$ 0.5 ‰ for the site Stumpy (Figs. 10 and 11 in Carlson et al., 2020). For $\delta^{13}C$ one reads 2.5‰ in Fig.3 in satisfactory agreement to the observation of values of about 2.2$\mp$0.5 ‰ in Fig. 11 in Carlson. For fast drip sites (Flatman) with $T_{drip}$ = 15 s variations of $\delta^{18,13}$ are significantly smaller of about 0.6 ‰. These variations however, are not seasonal. They exhibit heavier values at low $p_{CO2}$. A seasonal variation cannot be derived. Probably other processes may contribute to the variability of isotope composition at the Flatman site.

The reason for the smaller modeled difference in the isotopic composition of precipitated $CaCO_3$ under a shorter drip interval compared to a longer drip interval is obvious. For small drip times the amount of $HCO_3^-$ withdrawn from the $HCO_3^-$ pool is small and correspondingly also the increase in $\delta$. Therefore, the calcite deposited during this time until a new drop replaces the water layer on the stalagmite shows less increase in $\delta$ compared to a stalagmite with large drip times where the increase in the isotope composition of the $HCO_3^-$ pool is larger and consequently that in the calcite precipitated. For $T_{drip} > 4\tau_{prec}$ practically all calcite has been withdrawn from the solution and all isotopic imprints will not change anymore and become independent of drip time. The growth rate of the stalagmite, however, further decreases with $1/T_{drip}$. For such a situation, one must not correlate isotope imprints to growth rate (Stoll et al., 2015, Dreybrodt, 2016). To conclude this chapter, the observations of Carlson can be explained by slow outgassing of $CO_2$ driven by precipitation of calcite exclusively.

## 4. Prior calcite precipitation (PCP)

When water with calcium concentration of 2mmol/L, saturated with respect to calcite enters into the cave it may form a drop that directly falls to the apex of the stalagmite. We have discussed this condition so far. It is very likely, however, that the water flows down the cave wall forming a drip site at some protrusion from where it drips to the stalagmite. It may also flow down a stalactite where it forms a stalagmite below. Therefore the processes outgassing and precipitation of calcite occur before the water reaches the stalagmite.This way the water precipitates calcite during the time $T_{PCP}$ until it drips to the stalagmite (Fohlmeister et al., 2020). Because the depth of the water layer can change on the way to the drip site, precipitation times are unknown. However, the reduced concentration, $c_{Ca}(T_{PCP})$, of Ca at the drip site can be measured.

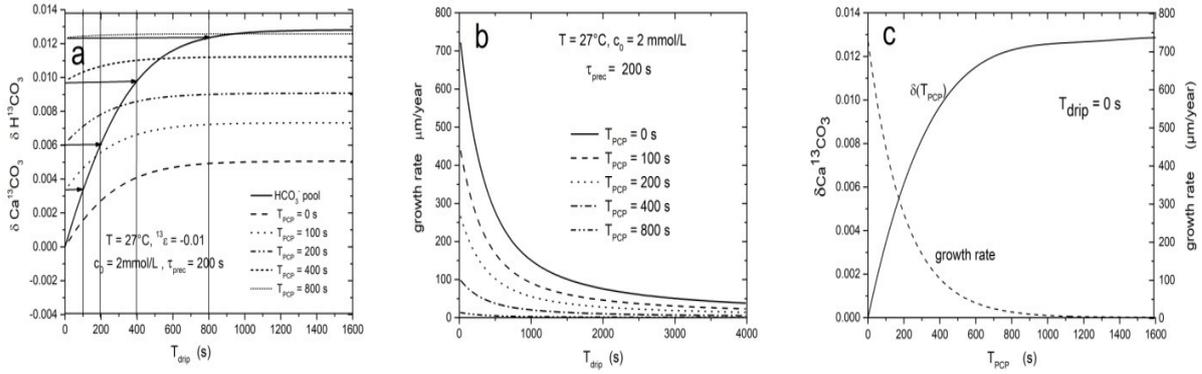

*Fig.4: δCa$^{13}$CO$_3$ in dependence on drip time T$_{drip}$ of the calcite deposited to the top of the stalagmite. a) The solid line depicts the isotope composition, δH$^{13}$CO$_3$, in the water during calcite precipitation after the solution got in contact with the cave atmosphere. Vertical lines represent the duration of PCP (T$_{PCP}$) of this drip water, after which the drop impinges on the stalagmite surface. Starting at this point carbonate precipitation contributes to the growth and isotopic composition of the speleothem according to the drip interval (T$_{drip}$). The isotopic composition of speleothem CaCO$_3$ is shown by the dashed and dotted lines in dependence of the drip interval after the solutions experienced a various duration of PCP. b) Growth rate of the stalagmite after PCP has been active for time T$_{PCP}$ in dependence on T$_{drip}$. c) δCa$^{13}$CO$_3$ of calcite (solid line) that is precipitated initially when the water has reached the stalagmite (T$_{drip}$ = 0) after it has experienced PCP during the time span T$_{PCP}$. The dashed line illustrates the corresponding growth rate See text.*

Because in a Rayleigh distillation process the increase of the isotope composition along the flow path is a function of $c_{ca}(T_{PCP})/c(0)$, the impact of PCP to the calcite deposited on the stalagmite can be calculated.

$$\delta(T_{PCP}) = \left(c_{Ca}(T_{PCP})/c_{Ca}^0\right)^{\mathcal{E}} \cdot (\delta(0)+1) - 1 \qquad (20)$$

δ(T$_{PCP}$) and δ(0) are the isotope compositions of HCO$_3$ in the water that drips to the stalagmite and that of the water entering the cave respectively. The solid line in Fig.4a depicts δH$^{13}$CO$_3$ of the HCO$_3^-$ pool in that water. For sake of simplicity we assume that the depth of the water film is, a = 0.01 cm during PCP and at the top of the stalagmite.

The isotope composition of CaCO$_3$ deposited on the stalagmite is given (Guo and Zhou, 2019) by eqn. 19 where c$_0$ is replaced by c(T$_{PCP}$) and by c(T$_{drip}$ + T$_{PCP}$).

$$\overline{R} = \frac{\int_{c_{Ca}(T_{PCP})}^{c_{Ca}(T_{drip}+T_{PCP})} R(c_{Ca}) \cdot dc_{Ca}}{\int_{c_{Ca}(T_{PCP})}^{c_{Ca}(T_{drip}+T_{PCP})} dc_{Ca}} + \varepsilon_{CaCO3/HCO3} = \frac{R(c_0) \int_{c_{Ca}(T_{PCP})}^{c_{Ca}(T_{drip}+T_{PCP})} (c_{Ca}/c_{Ca}^0)^\varepsilon \cdot dc_{Ca}}{\int_{c_{Ca}(T_{PCP})}^{c_{Ca}(T_{drip}+T_{PCP})} dc_{Ca}} + \varepsilon_{CaCO3/HCO3} =$$

$$= \frac{R(c_0) \int_{T_{PCP}}^{T_{drip}+T_{PCP}} \alpha \cdot (c_{Ca}(t)/c_{Ca}^0)^\varepsilon \cdot (c_{Ca}(t)-c_{Ca}^{eq}) \cdot dt}{\int_{T_{PCP}}^{T_{drip}+T_{PCP}} \alpha \cdot (c_{Ca}(t)-c_{Ca}^{eq}) \cdot dt} + \varepsilon_{CaCO3/HCO3}$$

(21)

PCP has a large impact to the isotope composition of the CaCO$_3$ on the speleothem (Fig. 4). The dependence of $\delta Ca^{13}CO_3$ on $T_{PCP}$ is shown in Fig. 4c. The full line depicts the isotope composition of the calcite deposited after PCP has been active during the time $T_{PCP}$. The values of $\delta Ca^{13}CO_3$ increase up to 13 ‰ for $T_{PCP} > 4\tau_{prec}$. This is a factor of two larger than what is possible by a change in drip time alone (Fig. 4a). The value of $T_{PCP}$ may change when the water supply to the drip site varies. In addition to this, variations of $\delta Ca^{13}CO_3$ are created by changes in $T_{drip}$. On the other hand, the initial growth rates decline rapidly with increasing $T_{PCP}$ as seen by the dashed line in Fig. 4c. Consequently, stalagmites experiencing PCP for time $T_{PCP} > 4\tau_{prec}$ exhibit growth rates less than 13 µm/year.

So the isotope composition of stalagmites with very low growth rates will likely show impact of PCP whereas those with large growth rates reflect probably drip sites with smaller times for $T_{PCP}$

The solid line in Fig.4a depicts $\delta H^{13}CO_3^-$ of the $HCO_3^-$ pool in the water. For sake of simplicity we assume that the depth of the water film is, a = 0.01 cm during PCP and at the top of the stalagmite and $\varepsilon_{CaCO3/HCO3}$ = 0. After time $T_{PCP}$ by Rayleigh distillation, the pool reaches the composition $\delta H^{13}CO_3$ shown by the arrows for the corresponding times. The isotope composition of the calcite $\delta Ca^{13}CO_3$ deposited at that moment is equal to $\delta H^{13,18}CO_3^-$ for very small drip time $T_{drip}$ = 0. For larger $T_{drip}$, $\delta Ca^{13,18}CO_3$ increases by Rayleigh distillation depicted by the corresponding dotted lines. Fig.4b shows the growth rate of the stalagmite in dependence of $T_{drip}$ for various values of $T_{PCP}$. After the time $T_{PCP}$ the water has needed to arrive at the drip site its Ca concentration has dropped exponentially with $\tau_{prec}$. Therefore, the precipitation rates at the surface of the stalagmite decline accordingly.

$\delta Ca^{13}CO_3$ in the calcite of stalagmites carries information on meteoric precipitation. Drip time $T_{drip}$ is correlated to the amount of meteoric precipitation. $T_{drip}$ increases with decreasing meteoric precipitation. As can be seen in Fig. 4a, one expects the most pronounced increase of $\delta Ca^{13, 18}CO_3$ when PCP is absent ($T_{PCP}$ = 0) and $T_{drip} > 4\tau_{prec}$. With increasing $T_{PCP}$ this value decreases. In parallel the growth rate of the stalagmite declines, see Fig. 4b. In conclusion, two coeval stalagmites, one without PCP and the other one with PCP active exhibit differing isotope imprints. Therefore, when selecting samples in the cave information on the drip site is valuable. Most favorable is a drip site where water drips directly from the cave ceiling thus preventing PCP. A drip site from a long stalactite likely will exhibit PCP whereas a short one may be suitable. For a stalactite, one can estimate $T_{PCP}$ by knowledge of its length, diameter and drip rate (Dreybrodt and Scholz, 2011). In general, one should accept samples only where $T_{PCP} < 0.2\tau_{prec}$. Recently Mickler et al. (2019) have observed PCP. They have measured temperature, pH, Ca, DIC-concentrations, and $\delta^{13}C_{DIC}$ of DIC in the drip water at a direct drip site where the water falls from a stalactite to flowstone and at an indirect drip site where it drips off after it has travelled a few meters downstream. At both drip sites more than 95% of the DIC was $HCO_3^-$ and the solution was supersaturated with $SI_{CaCO_3}$ of about 0.7. At the drip sites ISST Direct and Indirect from 10/19/2013 the Ca concentration decreased from 87 to 73 mg/L and $\delta^{13}C_{DIC}$ increased from -10.4 to -8.8. From this, using Rayleigh distillation $(\delta_{indir}+1000)/(\delta_{dir}+1000) = (c_{indir}/c_{dir})^\varepsilon$ they found $\varepsilon = -11$‰. This value is close to experimental findings in cave analogue lab experiments by Hansen et al. (2019). In these experiments supersaturated water drips to an inclined limestone plate (direct drip site) from where it flows down this plate and after a defined flow distance drops off the plate (indirect drip site).

**5. Impact of in cave effects by climatic conditions**

In this section we will show, in which way climate variations can be imprinted in speleothems, if we assume that the initial concentration and the initial isotopic composition remains the same for all cases considered. For illustration purposes, we also assume a well ventilated cave. During a cold climate conditions this cave has an average temperature of 5°C and atmospheric $p_{CO2}$ of 280 ppm prevails (Fig. 5, curve S1). Then the climate changed to 15°C and $p_{CO2}$ in the cave remained constant (Fig. 5, curve S2). After this, we assume an increasing atmospheric (and cave) $p_{CO2}$ to 450 ppm (curve S3), which caused an increase in temperature by 2°C. To find the isotope compositions of calcite during these stages we need to know the values of $c_{Ca}^{eq}$, $\tau_{prec} = a/\alpha$, and $^{13}\varepsilon$ or $^{18}\varepsilon$ respectively. For simplicity we assume the depth of the water, a = 0.01 cm, as constant for all three stages S. Furthermore, PCP is absent and the input concentration $c_{Ca}^o = 2 mmol/L$ is constant. The equilibrium concentrations, $c_{Ca}^{eq}$, are calculated for open system conditions using temperature, $p$, and $c_{Ca}^o$ as input parameters (Dreybrodt, 1988, program EQUILIBRIUM, updated by Franci Gabrovsek). We get the values of α in dependence of temperature from Baker et al. (1998). We use the values of ε from Hansen et al. (2019) as already stated. Table 2 lists all parameters. To calculate the isotope composition of the calcite deposited to the stalagmite for the three climate stages S1, S2, and S3 we use eqns. 18a and 19 with the parameters listed in Table 2. The climate conditions are mirrored by the parameters temperature T, $p_{CO2}$, $c_{Ca}^o$, and $T_{drip}$. $c_{Ca}^o$ depends on the $CO_2$ concentration in the soil and the ground air and may reflect vegetation. $T_{drip}$ is correlated to the amount of meteoric precipitation. Fig. 5 illustrates the result for $\delta^{13}C$ and $\delta^{18}O$ of precipitated $CaCO_3$ under various drip intervals. The large difference between the lines S1 and S2, S3 arises from the difference in temperature. Due to the significant temperature dependence of α, the precipitation time $\tau_{prec} = a/\alpha$ changes as well. For times t < 0.3·$\tau_{prec}$, the initial slope, m, of the curves is given (Dreybrodt and Scholz, 2011, Dreybrodt, 2019a) as

$$m = \varepsilon(1 - c_{Ca}^{eq}/c_{Ca}^{0})/\tau_{prec} \tag{22}$$

| stage S | $c_{eq}$ μmol/cm³ | α cm/s | $\tau_{prec}$ s | T °C | $p_{CO2}$ ppm |
|---|---|---|---|---|---|
| S1 | 0.64 | 4.7·10⁻⁵ | 1111 | 5 | 280 |
| S2 | 0.55 | 2.0·10⁻⁵ | 500 | 15 | 280 |
| S3 | 0.62 | 2.3·10⁻⁵ | 445 | 17 | 450 |

*Table 2: Values of $c_{eq}$, α, and $\tau_{prec}$ for climate states $S_1$, $S_2$, $S_3$, defined by temperature and $p_{CO2}$.*

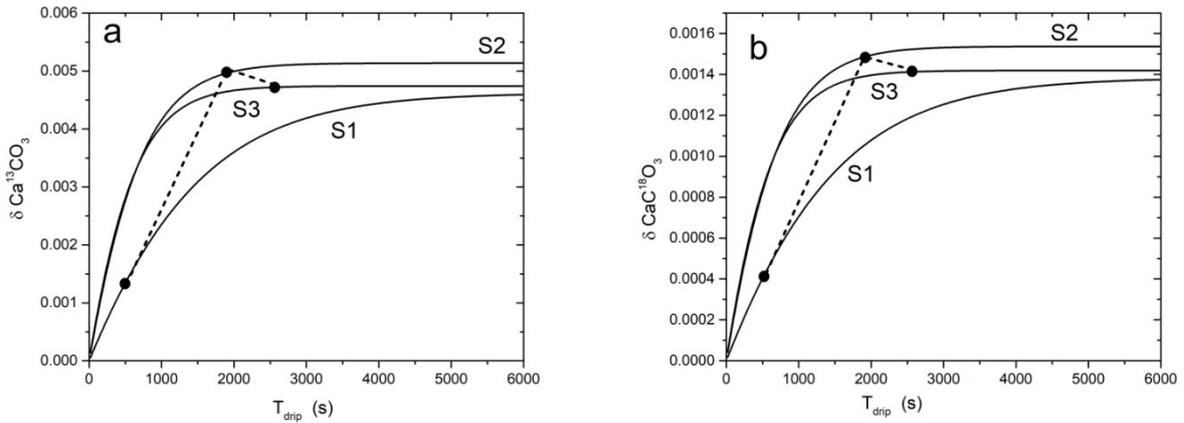

*Fig. 5. Isotope composition in dependence of drip time, $T_{drip}$, for different stages of climate. Each curve S is specified by cave temperature, cave $p_{CO2}$, and $c_{Ca}^o$. S1: $p_{CO2}$ = 280 ppm, $T_{cave}$ = 5°C, $c_{Ca}^o$ = 2mmol/L. S2: S1: $p_{CO2}$ = 280 ppm, $T_{cave}$ = 15°C, $c_{Ca}^o$ = 2mmol/L. S3: S3: $p_{CO2}$ = 450 ppm, $T_{cave}$ = 17°C, $c_{Ca}^o$ = 2mmol/L. Drip time, $T_{drip}$, is selected (black dots) as a measure of meteoric precipitation.*

The initial slope of S1 differs from those at higher temperatures, S2 and S3. This is a surprising result so far not discussed as reason of temperature imprint.

The final value of $\delta_{CaCO3}$ at large drip times $T_{drip} > 3\,\tau_{prec}$ is

$$\delta_{CaCO3}(3\tau_{prec}) = \left(c_{Ca}(3\tau_{prec})/c_{Ca}^{0}\right)^{\varepsilon} - 1 = \left(c_{Ca}^{eq}/c_{Ca}^{0}\right)^{\varepsilon} - 1 \tag{23}$$

$c_{Ca}^{eq} = K(T)\sqrt[3]{p_{CO2}}$ depends on $p_{CO2}$ and weakly on temperature. This way the isotope composition contains

information on CO₂ in the cave. Furthermore, the concentration of the water entering the cave may give information on climatic conditions.

The changes of $\delta_{CaCO3}$ can be read as follows. The black circles indicate the $\delta_{CaCO3}$ under conditions of climate, represented by curves S1, S2, and S3 and drip time represented by the black points. The drip times are related to the amount of meteoric precipitation. The shifts of $\delta_{CaCO3}$ by changing climate conditions are found by connecting two points. In our example the value of $\delta^{13}_{CaCO3}$ changes by 3.6‰ from S1 to S2 but only by 0.2‰ from S2 to S3 although T increased further. However, this T increase was overcompensated by the increase in cave air pCO2. The variations in $\delta^{18}_{CaCO3}$ are smaller by a factor of 3 due to the smaller $^{18}\varepsilon$ = -0.003 in comparison to $^{13}\varepsilon$ = -0.010.

An important result is the dependence on $T_{drip}$. Only stalagmites that grew with high water supply ( $T_{drip} \ll \tau_{prec}$) do not show imprints of in cave processes. They reflect the isotope composition of the water entering the cave. Such stalagmites have diameters larger than 10 cm ( Dreybrodt, 2008).Therefore, stalagmites with diameters above 10 cm are a good choice to be collected to avoid imprints of in cave effects.

## 6. Conclusion

We have discussed the processes of outgassing and precipitation of calcite from thin films of cave drip water and the evolution of its isotope composition. In short, Fig. 6 illustrates the results. It depicts the chemical pathway of the drip water as it enters into the cave, degasses, equilibrates, and precipitates calcite in a [$Ca^{2+}$] - pH - [$CO_2$] concentration coordinate system. The solution with high $CO_2$-concentration enters at point A. This water has just achieved $Ca^{2+}$saturation after dissolution of the carbonate host rock and $CO_2$ has not yet degassed and no carbonate has yet been precipitated. In other words, PCP has not been active. In the first step of diffusive outgassing (solid line) the $CO_2$-concentration drops but pH and Ca-concentration remain constant. The $CO_2$-pool is emptied without any chemical and isotopic effect to the $HCO_3^-$ pool. Therefore, this initial phase of the $CaCO_3$ precipitation process does not affect the isotope composition of the calcite to be precipitated later on.

After completion of outgassing (point B) with a characteristic time $\tau_{dif}$ the solution achieves equilibrium with respect to the $CO_2$-concentration of cave air (dashed line). The new equilibrium-pH is attained with time scale $\tau_{eq} \gg \tau_{dif}$. During equilibration between aqueous $CO_2$ and $H_2CO_3$ pH increases but there is yet no precipitation of calcite. Therefore, the $HCO_3^-$ pool remains closed. There is no effect to the isotope composition of the calcite that precipitates later on. At point C precipitation of calcite starts with **characteristic** time scales $\tau_{prec} \gg \tau_{eq}$ (dot-dashed line). The $CO_2$-concentration in the solution remains constant in equilibrium with the $p_{CO2}$ of the cave atmosphere. **For thin water depth as they occur on stalagmites there is no "gradient" of $p_{CO2}$ between gaseous and aqueous $CO_2$ during this time.** During precipitation of CaCO$_3$, further outgassing of CO$_2$ is driven by precipitation of calcite. Precipitation stops when the Ca-concentration reaches $c_{Ca}^{eq} = K(T)\sqrt[3]{p_{CO2}}$ in equilibrium with the p$_{CO2}$ of the cave atmosphere. The HCO$_3^-$ pool remains unaffected during the first two steps (Fig. 6; solid and dashed lines). Both, its concentration and its isotope composition stay constant. Only during precipitation of calcite (Fig. 6, dot-dashed line) CO$_2$, stemming from the HCO$_3^-$ pool, is released into the atmosphere leading to changes in the isotopic composition of this pool. The $\delta^{13}$C and $\delta^{18}$O values of HCO$_3^-$ increase in the water and accordingly also in the precipitated calcite.

In addition, we elaborated on the precipitated calcite, which forms speleothems. While our approach agrees well with earlier ones with respect to the isotopic evolution of HCO$_3^-$ on drip rate and p$_{CO2}$ changes, we discuss the isotopic evolution of CaCO$_3$ as well. We show, that drip rate changes are only important as long as drip rate is smaller than T$_{drip}$ < 2·$\tau_{prec}$.

Drip rate changes on longer time scales will barely affect the isotopic composition of precipitated calcite as more than 86% of precipitation occurs in this first time interval. The same is true for precipitated CaCO$_3$. Thus, observations of strong changes in growth rates and isotopic composition are barely explainable by drip rate changes alone.

We focused as well on the PCP process, which we investigated with respect to growth rate and the stable carbon and oxygen isotopic composition. We show that shifts often observed in the isotopic composition of CaCO$_3$ and PCP can explain the corresponding change of growth rate. In comparison to drip rate changes, changes in the time for PCP have a more important influence on the final isotopic composition of precipitated CaCO$_3$ and the growth rate of speleothems. One has to consider this in any speleothem climate study. It will be very important to quantify the degree of PCP for climate reconstructions. Future efforts should focus on this task. Finally, we have discussed the contribution to isotope composition by in cave effects under various climate conditions.

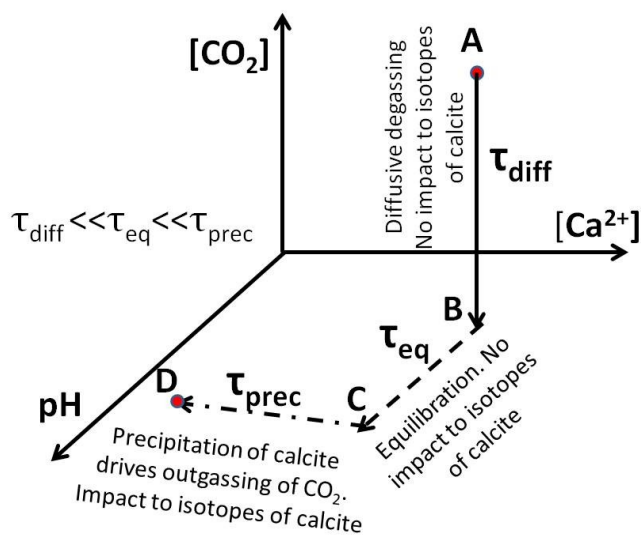

*Fig. 6: The pathway of chemical evolution of a CaCO$_3$-CO$_2$ aqueous solution after dripping to a stalagmite. Step 1 (solid): diffusive degassing of aqueous CO$_2$. Step 2 (dashed): equilibration to the low p$_{CO2}$. Step 3 (dot-dash): precipitation of calcite and decrease of pH.*

In summary, we have presented a method that allows calculation of the isotope composition $\delta_{CaCO3}$ in its dependence on temperature, $p_{CO2}$ in the cave, and the concentration $c_{Ca}^o$ of the water entering the cave. We consider also the effects of PCP. This enables us to explore isotope compositions at various stages of climate conditions.

Our findings have important implications to understand precipitation of travertine in karst springs. In contrast to precipitation of calcite from a physically well constrained stagnant thin layer of solution precipitation of calcite in fast turbulently flowing streams is not well constrained for the following reasons. The first step of outgassing of molecular $CO_2$ is described by a Two-Layer model of the gas-liquid interface ( Liss et al.,1974). In this model the bulk bodies of water and gas are well mixed by turbulence. Gas transport proceeds through gas and liquid phase interfacial boundary layers separating the liquid-gas phase across which the gas is transported by molecular diffusion. The depths of the boundary layers depend on the flow velocity of the water. From this model a transport constant k (m/s) can be derived. The time $\tau_{diff}$ for outgassing is given by $\tau_{diff}$ = D/k where D is the depth of the water. k depends on the flow velocity of the water. The time $\tau_{diff}$ in nature is on the order of several minutes. This value can be estimated from the distance the water flows from the spring outlet until precipitation of calcite becomes active (Usdowski et al., 1979, Dreybrodt et al.,1992, Liu et al.,1995). If $\tau_{diff} \gg \tau_{eq}$ during the entire time of outgassing aqueous $CO_2$ and the $HCO_3^-$ pool are in chemical equilibrium. Since precipitation of calcite is not active until SI has reached a critical value the $HCO_3^-$ pool remains isolated and its isotope composition is not altered. As long as calcite precipitation is absent this is valid generally for all values of $\tau_{diff}$. This is in agreement to recent findings of Yan et al., 2020. In conclusion our findings on the first step of diffusive outgassing in thin layers of stagnant solution are generally valid also for turbulent flow in travertine depositing streams

For the precipitation driven outgassing of $CO_2$ the hydrodynamics of flow play an important role. The turbulently well mixed bulk of the calcite depositing solution is separated by a diffusion boundary layer (DBL) from the solid calcite surface (Dreybrodt and Buhmann, 1991, Liu and Dreybrodt, 1997). Precipitation rates are determined by the thickness of the DBL. The rate equation R = $\alpha(c-c_{eq})$ remains valid, but $\alpha$ depends on the depth of the DBL. Therefore, in contrast to precipitation in a stagnant water layer on top of a stalagmite, depending on the flow conditions in the travertine depositing stream $\alpha$ changes along its flow path (Dreybrodt et al., 1992, Liu et al., 1995). Since flow conditions change seasonally the isotope compositions at each location in the stream change accordingly (Liu et al. 2010). These complications have to be taken into account for interpretation of paleoclimate signals. Further complications arise from observations that the fractionation constants depend on hydrodynamic conditions as found for $\varepsilon_{HCO3\text{-}CaCO3}$ in a pool and its water flowing in rapid flow across the rim of the dam confining the pool (Yan et al., 2021).

We hope that our work may give help to gain insights to the interpretation of isotope compositions of travertine.

**Acknowledgements**

We thank two anonymous reviewers for their comments, which helped to improve the manuscript. We highly appreciate the editorial expertise of Michael Böttcher.

**Appendix**

**A1. Equilibration**

Here we give a comprehensive description of the equilibration processes. After completion of diffusive outgassing, the concentration of aqueous $CO_2$ has dropped to a low value. Therefore, the concentration of carbonic acid, $H_2CO_3$, must adjust by the reaction $H_2CO_3 \leftrightarrow CO_2 \uparrow + H_2O$. This reaction is slow in comparison to the initial $CO_2$ degassing, i.e., the characteristic time constant $\tau_{eq}$ is more than one order of magnitude larger than $\tau_{diff}$ (Zeebe and Wolf-Gladrow, 1999, Dreybrodt and Scholz, 2011). Although the concentration of carbonic acid is only about 0.2% of the concentration of aqueous $CO_2$ it plays a dominant role in the carbonate chemistry.

The $CO_2$ created escapes from the solution by diffusive outgassing with time scale $\tau_{dif} \ll \tau_{eq}$. The concentrations of $^{13}C^{16}O^{16}O$ and $^{12}C^{18}O^{16}O$ in the aqueous $CO_2$-pool are fixed to their concentration in the cave atmosphere by Henry's law. Therefore, neither its concentration nor its isotope composition does change during equilibration of $H_2CO_3$ with $CO_2$. Simultaneously equilibrium between $HCO_3^-$ and $H_2CO_3$ is established instantaneously by the fast reaction $H^+ + HCO_3^- \rightarrow H_2CO_3$. In this reaction, the concentrations of $H^+$ and $HCO_3^-$ decrease. At all times during equilibration of $H_2CO_3$ and $CO_2$, $HCO_3^-$ and $H_2CO_3$ are in equilibrium. Therefore mass action law

$$\frac{[H^+]\cdot[HCO_3^-]}{[H_2CO_3]}=K \tag{A.1}$$

is valid at all times.

Immediately after out gassing and before equilibration the initial concentration $[H_2CO_3^{in}]$ is high in equilibrium with the high concentration $[CO_2]^{high}$ in the drip water. $[H_2CO_3^{in}]=K_{CO2}\cdot[CO_2]^{high}$. After equilibration, we have $[H_2CO_3^{eq}]=K_{CO2}\cdot[CO_2]^{low}$. Where $[CO_2]^{low}$ is the concentration after outgassing. At the onset of equilibration one has

$$\frac{[H^+]^{in}\cdot[HCO_3^-]^{in}}{[CO_2]^{high}}=K\cdot K_{CO2} \tag{A.2}$$

After equilibration, $[H^+]^{eq} = [H^+]^{in} - x$ and $[HCO_3^+]^{eq} = [HCO_3^+]^{in} - x$. x is the loss in the H$^+$ and HCO$_3^-$ concentration respectively. Furthermore, $[CO_2]^{low} = y \cdot [CO_2]^{high}$  $y \ll 1$. Using eqn. A2 one gets

$$\frac{[H^+]^{eq} \cdot [HCO_3^-]^{eq}}{[CO_2]^{low}} = \frac{([H^+]^{in} - x) \cdot ([HCO_3^+]^{in} - x)}{y \cdot [CO_2]^{high}} = \frac{[H^+]^{in} \cdot [HCO_3^-]^{in}}{[CO_2]^{high}} \quad (A.3)$$

This is a quadratic equation in x. Regarding $[H^+]^{in} \ll [HCO_3^-]^{in}$ its solution is

$$x = [H^+]^{in} \cdot (1 - y) \cong [H^+]^{in} \quad (A.4)$$

At pH = 7, $[H^+]^{in}$ is $10^{-4}$ mmol/L.

Therefore the change of ca. $10^{-4}$ mmol/l in the HCO$_3^-$ concentration of about 4 mmol/L during equilibration can be safely neglected and the system can be regarded as closed with respect to HCO$_3^-$ and also as shown already with respect to CO$_2$.

This can be summarized by simple reasoning. During the time of isotopic and chemical equilibration between dissolved CO$_2$ and HCO$_3^-$ (via H$_2$CO$_3$) until start of precipitation of calcite the change of HCO$_3^-$ concentration by the instantaneous reaction H$^+$ + HCO$_3^-$ →H$_2$CO$_3$ is equal to the change of the H$^+$ concentration of about $10^{-7}$ mol/L .

**A2. Classical and extended Rayleigh equation**

The theoretical basis for Rayleigh distillation is chemical and isotope equilibrium of all reacting agents. The Classical Rayleigh Equation is

$$R_{classical}(c_{Ca}(t)) = \left(c_{Ca}(t)/(c_{Ca}^0)\right)^\varepsilon = 1 + \delta_{CaCO3}(t) \quad (A.5)$$

Alternatively, a kinetic approach has been suggested (Dreybrodt, 2008, Dreybrodt and Scholz, 2011, Dreybrodt, 2016). Here, the isotopologues react independently of each other. In the reaction, $F = \alpha(c_{Ca} - c_{Ca}^{eq})$ the reaction constant, α, differs for the heavy and the light isotopologues. The isotope fractionation $\alpha^{kin}$ is defined by the ratio $\alpha_{heavy}/\alpha_{light} < 1$. Furthermore, one accounts for different equilibrium concentrations with respect to calcite of the light and heavy isotopes by a fractionation factor, $\gamma \cong 1$. From this, one obtains the Extended Rayleigh equation.

$$R_{extended}(c_{Ca}(t)) = \left[\left(\frac{c_{Ca}(t) - c_{Ca}^{eq}}{c_{Ca}^0 - c_{Ca}^{eq}}\right)^{\alpha_{kin}} \cdot \left(1 - \gamma \frac{c_{Ca}^{eq}}{c_{Ca}^0}\right) + \gamma \frac{c_{Ca}^{eq}}{c_{Ca}^0}\right] \cdot \frac{c_{Ca}^0}{c_{Ca}(t)} = 1 + \delta_{CaCO3}^{extended}(t) \quad (A.6)$$

As we are interested only in the change of the isotope composition, we take $R_{extended}(c_{Ca}^0) = 1$

Integrating for each equation with the fractionation constants as given in the main text using

$$\bar{R} = \frac{\int_{c_{Ca}(0)}^{c_{Ca}(Tdrip)} R(c_{Ca}) \cdot dc_{Ca}}{\int_{c_{Ca}(0)}^{c_{Ca}(Tdrip)} dc_{Ca}} \quad (A.7)$$

yields $\delta_{CaCO3}^{extended}(t)$ and $\delta_{CaCO3}^{classical}(t)$. These for comparison are depicted in Fig. A1.

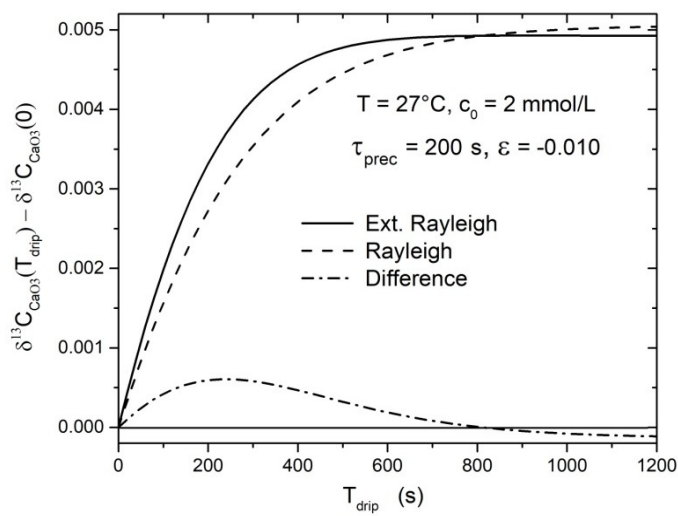

*Fig. A1 : $\delta_{CaCO3}^{extended}(T_{drip})$ and $\delta_{CaCO3}^{classical}(T_{drip})$ in comparison. The lowest line shows their difference.*